\documentstyle[epsfig]{basi}
\def\bi{\bibitem{}}

\def\ni{\noindent}
\def\beb{\begin{thebibliography}{10}}
\def\eeb{\end{thebibliography}}
\def\bei{\begin{itemize}}
\def\eei{\end{itemize}}
\def\bef{\begin{figure}}
\def\eef{\end{figure}}
\def\ben{\begin{enumerate}}
\def\een{\end{enumerate}}
\def\beq{\begin{equation}}
\def\eeq{\end{equation}}
\def\ber{\begin{eqnarray}}
\def\eer{\end{eqnarray}}
\def\edo{\end{document}}

\begin{document}
\title[neutron star field evolution]{Diamagnetic Screening of the Magnetic Field in Accreting Neutron Stars}
\author[Konar and Choudhuri]%
       {Sushan Konar$^1$, Arnab Rai Choudhuri$^2$\thanks{e-mail:sushan@iucaa.ernet.in,arnab@physics.iisc.ernet.in} \\ 
        $^1$Inter-University Centre for Astronomy \& Astrophysics, Pune 411007 \\
        $^2$Department of Physics, Indian Institute of Science, Bangalore 560012}
\maketitle
\label{firstpage}
\begin{abstract}
A possible mechanism for screening of the surface magnetic field of an accreting neutron star, by the accreted material, 
is investigated. In particular, we investigate the nature of the evolution of the internal field configuration in the case
of a) a polar cap accretion and b) a spherical accretion.
\end{abstract}


\ni Observations of radio pulsars and X-Ray binaries imply that the evolution of magnetic field is related to mass accretion 
onto a neutron star (Bhattacharya \& Srinivasan 1991). Theoretical efforts in modeling the plausible mechanisms have so far 
been concentrated on the ohmic dissipation of crustal currents (Konar \& Bhattacharya 2001). One of the novel mechanisms which 
could also be operative, in an accreting neutron star, is the diamagnetic screening of magnetic field by the accreting material 
(Romani 1990).

\ni Recently, Choudhuri \& Konar (2002) have done detailed modeling of the flow of the accreting material in the surface 
layers of a neutron star. The accreted material, confined to the poles by strong magnetic stresses, accumulates in a column 
and sinks below the surface when the pressure of the accretion column exceeds the magnetic pressure. From the bottom of the 
accretion column, the material moves to the lower latitudes in an {\em equator-ward} flow. Material from both the poles 
meet at the equator and submerges, pushing against the solid interior and displacing it very slowly in a {\em counter-flow} 
downwards as well as to higher latitudes. And, in the very deep layers the material moves radially inwards due to overall 
compression of the star. We have modelled the flow ensuring that it has all the above characteristics and satisfies the 
continuity equation (incompressible matter). In Fig.1 we show the profile of this velocity flow (a) for a low rate of 
accretion (polar cap accretion) and, (b) for a higher rate of accretion, (more towards spherical accretion).          

\ni Now, this accreted material, flowing horizontally from the polar caps, drags the magnetic field lines with it because 
of 'flux-freezing' resulting in the distortion of the original field structure. This distortion gives rise to an effective 
screening of the observable field over a time-scale of the material flow. Assuming an initial central dipolar configuration
for the field (Fig.2a) we show the final field configurations (obtained by solving the induction equation in 2-D) in Fig.2b 
\& Fig.2c corresponding to the two velocity profiles of Fig.1. Details of these calculations can be found in
Konar \& Choudhuri (2002).

\bef
\begin{center}{
\epsfig{file=fig1a.ps,width=80pt}
\hspace{0.75cm}
\epsfig{file=fig1b.ps,width=80pt}}
\end{center}
\caption[]{Flow velocity profiles in the region $0 \leq \theta \leq \pi/2$ and $0.25 \leq r \leq 1.0$, for -
{\bf a)} a polar cap accretion, {\bf b)} a more spherical accretion.}
\label{fig1}
\eef          

\bef
\begin{center}{
\epsfig{file=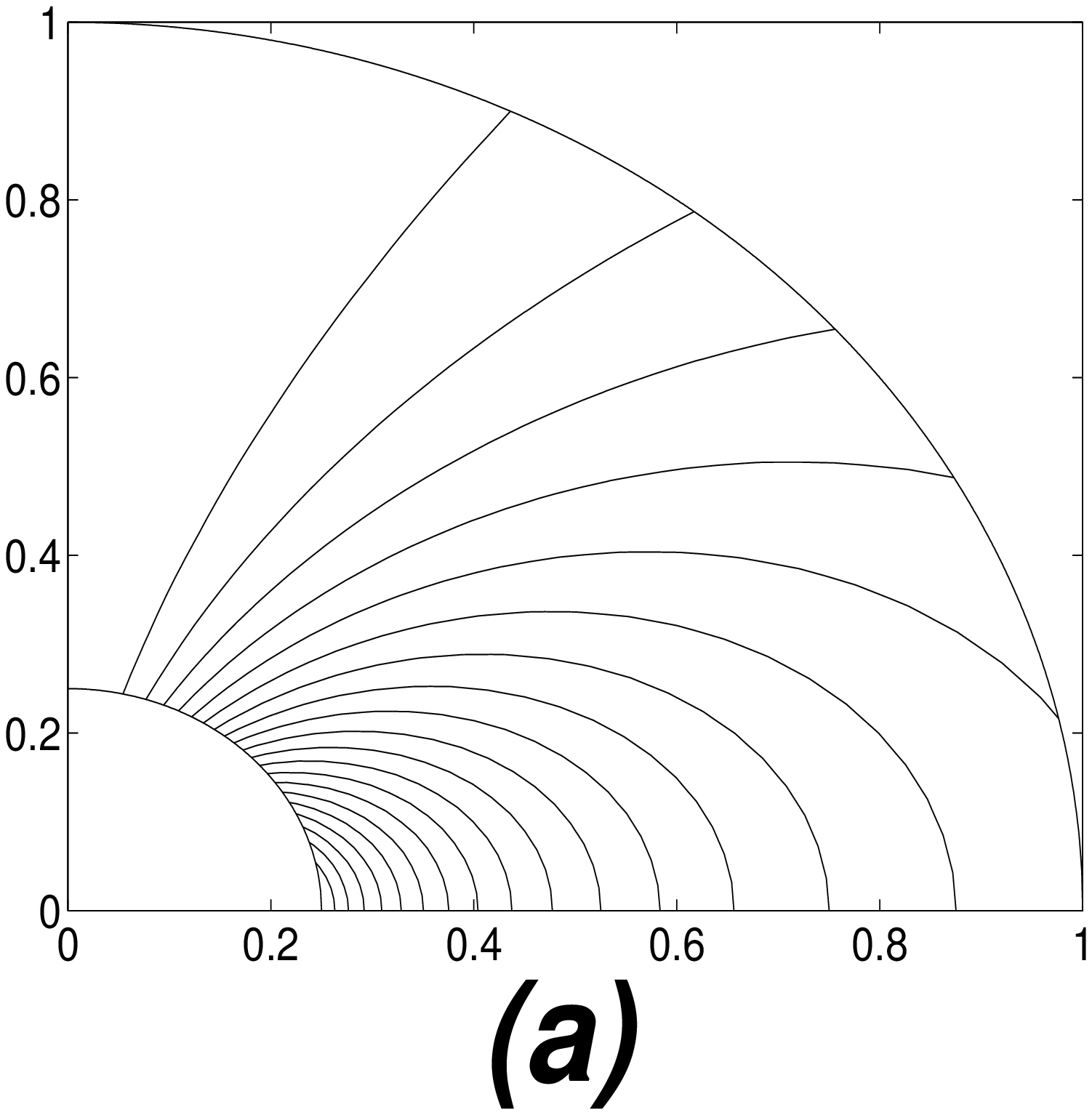,width=100pt}
\hspace{0.25cm}
\epsfig{file=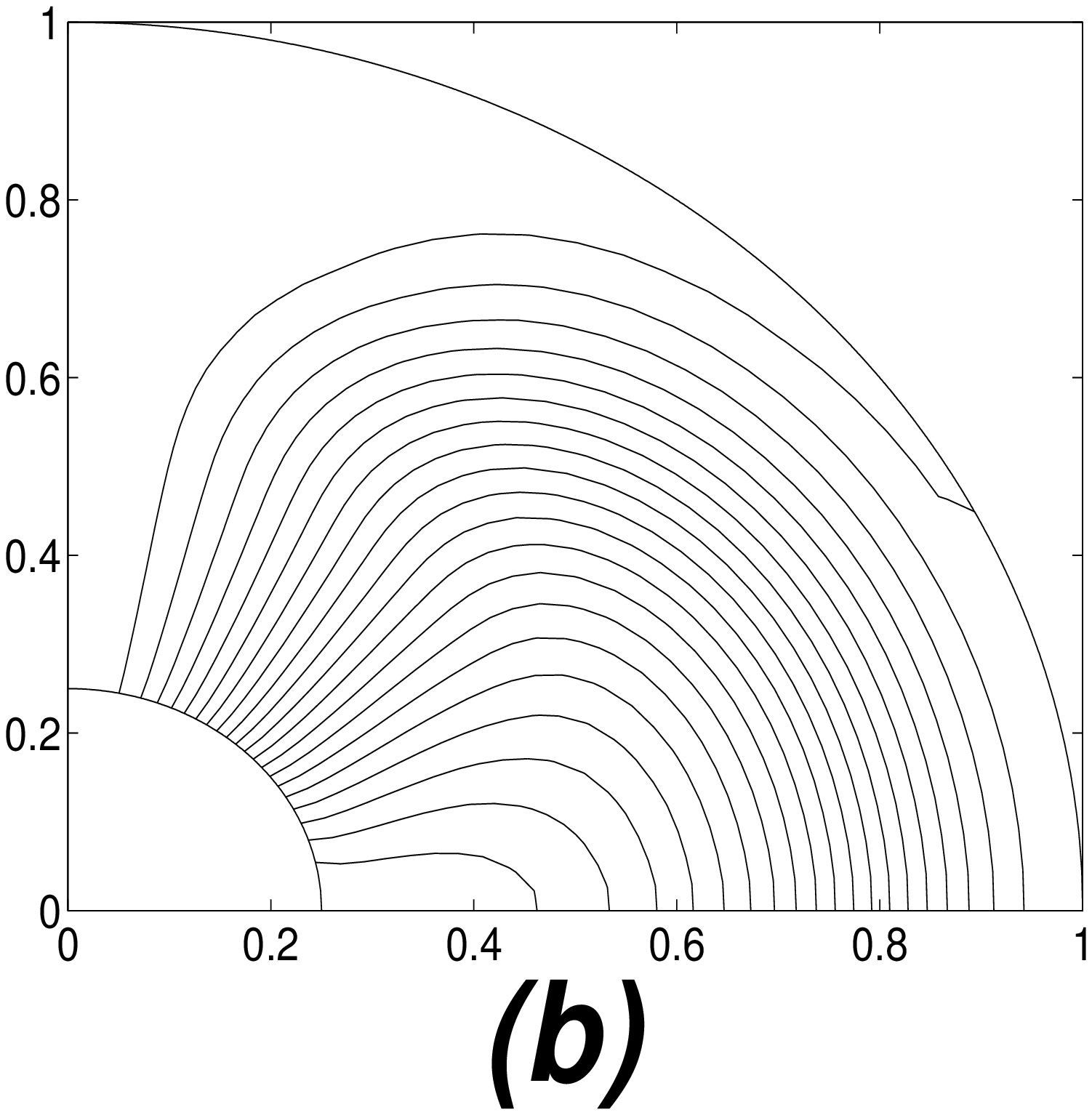,width=100pt}
\hspace{0.25cm}
\epsfig{file=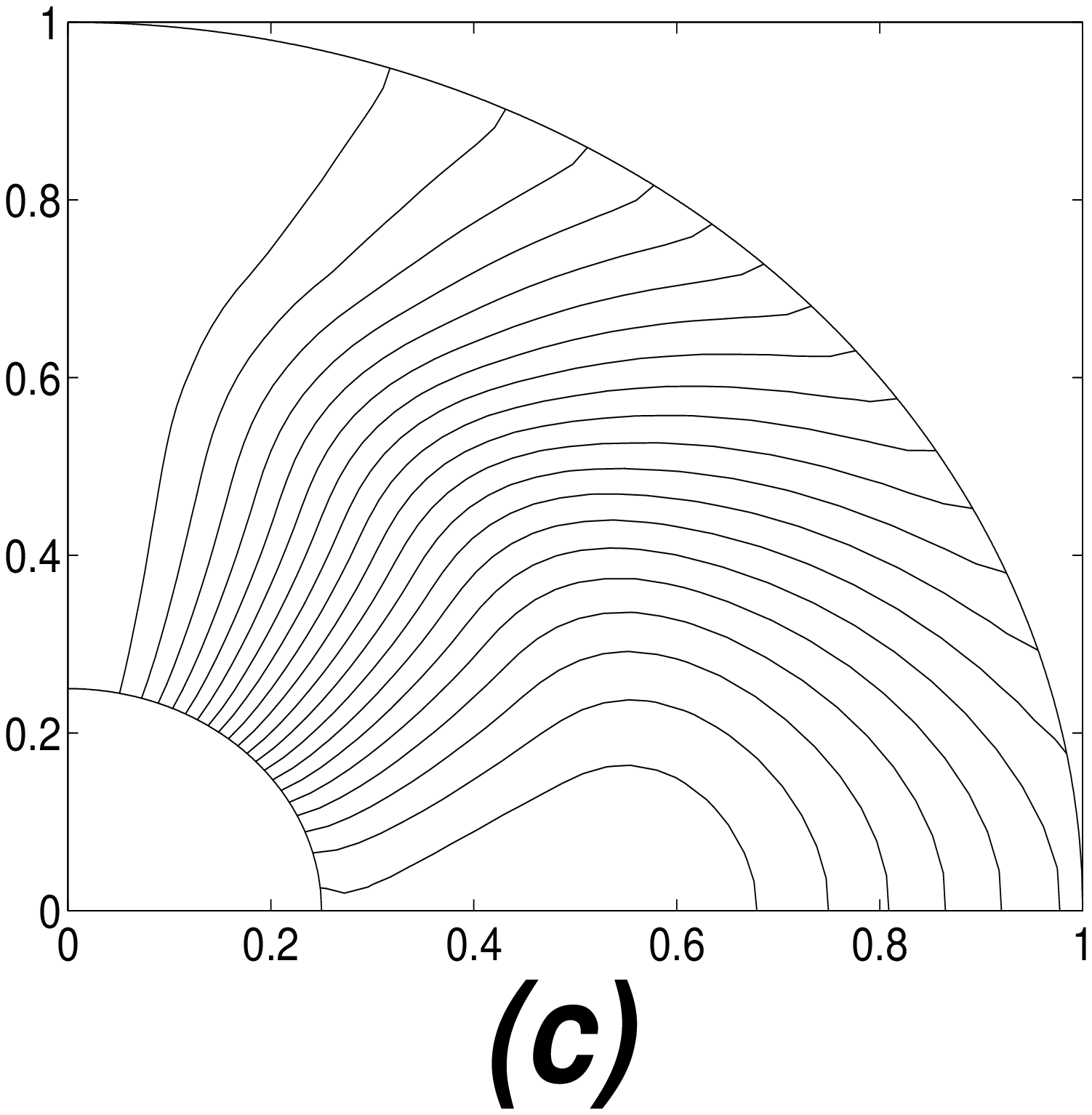,width=100pt}}
\end{center}
\caption[]{{\bf a)} Initial field configuration, assuming a central dipole. {\bf b)} \& {\bf c)} are evolved 
configurations corresponding to velocity profiles of {\bf Fig.1a} and {\bf Fig.1b}.}
\label{fig2}
\eef          

\beb
\bi Bhattacharya, D. and Srinivasan, G., 1991, \newblock In Ventura, J. and Pines, D., editors,
    {\it Neutron Stars: Theory and Observation}, page~219, Kluwer Academic Publishers, Dordrecht.
\bi Choudhuri, A.~R. and Konar, S., 2002, \newblock{\it MNRAS}, 332, 933
\bi Konar, S. and Bhattacharya D., 2001, \newblock In Kouveliotou, C. et al, editors, 
    {\it The Neutron Star Black Hole Connection}, page~71, Kluwer Academic Publishers, Dordrecht. 
\bi Konar, S. and Choudhuri, A.~R., 2002, in preparation
\bi Romani R. ~W., 1990, Nat, 347, 741
\eeb
\label{lastpage}
\end{document}